\documentclass[english]{article}
\usepackage[T1]{fontenc}
\usepackage[latin9]{inputenc}
\usepackage{units}
\usepackage{amstext}
\usepackage[all]{xy}

\makeatletter
\@ifundefined{date}{}{\date{}}
\usepackage[all]{xy}
\xyoption{rotate}
\usepackage{graphicx}
\usepackage{cite}

\makeatother

\usepackage{babel}
\begin{document}
\title{Contextuality and Random Variables}
\author{Ehtibar Dzhafarov}
\maketitle
\begin{abstract}
Contextuality is a property of systems of random variables. The identity
of a random variable in a system is determined by its joint distribution
with all other random variables in the same context. When context
changes, a variable measuring some property is instantly replaced
by another random variable measuring the same property, or instantly
disappears if this property is not measured in the new context. This
replacement/disappearance requires no action, signaling, or disturbance,
although it does not exclude them. The difference between two random
variables measuring the same property in different contexts is measured
by their maximal coupling, and the system is noncontextual if one
of its overall couplings has these maximal couplings as its marginals. 
\end{abstract}

\section{Preamble}

Quantum physicists like telling people (perhaps even themselves) that
their field is strange and counterintuitive. Contextuality, especially
when it takes the form of nonlocality, is one of its strange and counterintuitive
notions. Indications of contextuality, such as violations of the Bell-type
inequalities, are sometimes referred to as paradoxes. Like everything
else in quantum physics, contextuality involves probabilities, hence
random variables. And these, unlike the physical issues described
by them, are usually taken to be clear and well-known: nothing strange
or counterintuitive about random variables, they are merely mathematical
tools, on a par with derivatives and integrals. 

However, if I had any propensity to mystify my readers, I would argue
that random variables are very strange objects. A random variable
is a pure potentiality until it is realized, i.e., until it ``collapses''
into a single value being observed. Why is it less intriguing that
the measurement problem in quantum physics (the wonderment at why
the Schr\"odinger wave, which is essentially a special way of describing
a random variable with a spatiotemporal distribution, collapses into
a specific value being observed)? The textbook definition says that
a random variable is a function from a probability space to a measurable
space, but one would look in vain for any utilization of this fact
in the quantum physical literature. What are these probability spaces
on which the random variables are defined? How does one know that
two observations belong to a single random variable rather than two
different variables? Can one speak of values of a random variable
counterfactually, in terms of what its value might have been had it
not been what it was observed to be? Questions and wonderments like
this can be multiplied. I am not, however, into mystifying my readers.
All these questions have clear answers, but this clarity is not of
an evident variety, one cannot achieve it without nontrivial conceptual
work. And once one achieves clarity about random variables, I will
argue, this clarity is imparted on the substantive issues they describe,
contextuality including. 

One researcher who forcefully argued that contextuality and nonlocality
are primarily matters of probability theory rather than physics was
Andrei Khrennikov \cite{Khrennikov2008,Khrennikov2009}. He does not
seem to maintain this position currently, and his arguments when he
maintained it were different from those presented in this paper. Nevertheless,
I think my views are close to Khrennikov's former views in spirit.

\section{Random variables within a system}

In 1989 David Mermin published a popular-level discussion of the nonlocal
form of contextuality \cite{Mermin1989} (based on his 1981 work,
added to \cite{Mermin1989} as an appendix). I will present Mermin's
reasoning in a modified form. Consider the well-known Alice-Bob scenario,
in which Alice chooses between two settings, denoted 1 and 2, and
Bob chooses between his two settings, also denoted 1 and 2. The outcomes
of Alice's measurements at either setting can be $+1$ or $-1$, and
the same is true for Bob's measurements. Because of the way the experiment
is set up (e.g., with Alice's and Bob's measurements being spacelike
separated), Alice's choice of a setting cannot influence Bob's measurements,
and vice versa. Let us use the term ``context $i$$j$'' to describe
the situation in which Alice chooses setting $i\in\left\{ 1,2\right\} $
and Bob chooses setting $j\in\left\{ 1,2\right\} $. Suppose that
the outcomes of the two measurements in the four contexts have the
following probabilities: 
\begin{equation}
\begin{array}{c}
\begin{array}{c|c|c|c}
\textnormal{context 22} & \begin{array}{c}
\textnormal{Bob's}\\
\textnormal{outcome}
\end{array}=+1 & \begin{array}{c}
\textnormal{Bob's}\\
\textnormal{outcome}
\end{array}=-1\\
\hline \textnormal{Alice's outcome}=+1 & 0 & \nicefrac{1}{2} & \nicefrac{1}{2}\\
\hline \textnormal{Alice's outcome}=-1 & \nicefrac{1}{2} & 0 & \nicefrac{1}{2}\\
\hline  & \nicefrac{1}{2} & \nicefrac{1}{2}
\end{array}\\
\\
\begin{array}{c|c|c|c}
\textnormal{contexts 11, 12, 21} & \begin{array}{c}
\textnormal{Bob's}\\
\textnormal{outcome}
\end{array}=+1 & \begin{array}{c}
\textnormal{Bob's}\\
\textnormal{outcome}
\end{array}=-1\\
\hline \textnormal{Alice's outcome}=+1 & \nicefrac{1}{2} & 0 & \nicefrac{1}{2}\\
\hline \textnormal{Alice's outcome}=-1 & 0 & \nicefrac{1}{2} & \nicefrac{1}{2}\\
\hline  & \nicefrac{1}{2} & \nicefrac{1}{2}
\end{array}
\end{array}.\label{eq:distributions}
\end{equation}
This describes what is commonly referred to as a \emph{PR-box} \cite{PR1994},
a highly contextual system by all reasonable measures \cite{KujDzh2019},
and one that violates the CHSH inequalities \cite{CHSH1969} to the
maximal extent algebraically possible (I will explain this in Section
3). Mermin's reasoning is aimed at showing that there is something
paradoxical (``extremely perplexing,'' he says) about such a system
of random variables.\footnote{\label{fn:Mermin-does-mention}Mermin does mention the distributions
(\ref{eq:distributions}), with perfect correlations and anticorrelations,
but for a detailed reasoning he uses the distributions in which the
joint probabilities $\nicefrac{1}{2}$ and $0$ are replaced with
$(1+\cos\frac{\pi}{4})/4$ and $(1-\cos\frac{\pi}{4})/4$, respectively.
The difference is not significant for my presentation.}

We begin with context 11, and denote the two random variables representing
the outcomes of Alice's and Bob's measurements as follows:

\begin{equation}
\begin{array}{cc}
\textnormal{context }11 & \left[\begin{array}{r|c}
\textnormal{Alice's setting 1} & \textnormal{random variable }A_{1}\\
\\
\textnormal{Bob's setting 1} & \textnormal{random variable }B_{1}
\end{array}\right.\Pr\left[A_{1}=B_{1}\right]=1.\end{array}\label{eq:context 11}
\end{equation}
That the probability of $A_{1}=B_{1}$ is 1 follows from the joint
distribution (\ref{eq:distributions}) for context 11. 

Now, Mermin proposes what he calls the \emph{Strong Baseball Principle}
(SBP),\footnote{So dubbed because of the simile with the belief that watching a baseball
game on one's TV cannot affect the game's outcome.} that I will formulate as follows: 
\begin{description}
\item [{(SBP)}] if Alice's random variable at her setting $i$ can in no
way be influenced by Bob's choice of his setting, then the same random
variable can represent her measurement outcomes in both context $i1$
and context $i2$; analogously, Bob's measurement outcomes at his
setting $j$ can be represented by the same random variable in both
context $1j$ and context $2j$. 
\end{description}
In particular, in context $12$, Alice's measurement outcomes can
be represented by the same $A_{1}$ as they are in context 11, and
in context 21, Bob's measurement outcomes can be represented by the
same $B_{1}$ as they are in context 11:
\begin{equation}
\begin{array}{ccc}
\textnormal{context }12 & \left[\begin{array}{r|c}
\textnormal{\textnormal{Bob's setting }2} & ?\\
\textnormal{Alice's setting }1 & A_{1}
\end{array}\right.\\
\\
\textnormal{context }11 & \left[\begin{array}{r|c}
\textnormal{Alice's setting 1} & A_{1}\\
\\
\textnormal{Bob's setting 1} & B_{1}
\end{array}\right. & \Pr\left[A_{1}=B_{1}\right]=1\\
\\
\textnormal{context }21 & \left[\begin{array}{r|c}
\textnormal{Bob's setting }1 & B_{1}\\
\textnormal{\textnormal{Alice's setting }2} & ?
\end{array}\right.
\end{array}.
\end{equation}
We can easily fill in the places held by the question marks. We know
from (\ref{eq:distributions}) that the two random variables in context
12 are perfectly correlated, so once we have determined one of them,
the other must copy it (and the same holds for context 21):
\begin{equation}
\begin{array}{cc}
\textnormal{context }12 & \left[\begin{array}{r|cc}
\textnormal{\textnormal{Bob's setting }2} & B_{2}\\
 &  & \Pr\left[A_{1}=B_{2}\right]=1\\
\textnormal{Alice's setting }1 & A_{1}
\end{array}\right.\\
\\
\textnormal{context }11 & \left[\begin{array}{r|cc}
\textnormal{Alice's setting 1} & A_{1}\\
 &  & \Pr\left[A_{1}=B_{1}\right]=1\\
\textnormal{Bob's setting 1} & B_{1}
\end{array}\right.\\
\\
\textnormal{context }21 & \left[\begin{array}{r|cc}
\textnormal{Bob's setting }1 & B_{1}\\
 &  & \Pr\left[A_{2}=B_{1}\right]=1\\
\textnormal{\textnormal{Alice's setting }2} & A_{2}
\end{array}\right.
\end{array}.\label{eq:transplanting1}
\end{equation}
Then we apply SBP once again, and conclude that in context 22 the
measurements by Alice and Bob can be represented by the same random
variables $A_{2}$ and $B_{2}$ as they are in contexts 21 and 12,
respectively:
\begin{equation}
\begin{array}{cc}
\textnormal{context }22 & \left[\begin{array}{r|c}
\textnormal{Alice's setting 2} & A_{2}\\
\\
\textnormal{Bob's setting 2} & B_{2}
\end{array}\right.\Pr\left[A_{2}=B_{2}\right]=1.\end{array}\label{eq:transplanting2}
\end{equation}
That the probability of $A_{2}=B_{2}$ is 1 follows from the chain
\begin{equation}
\Pr\left[A_{2}=B_{1}\right]=1,\Pr\left[B_{1}=A_{1}\right]=1,\Pr\left[A_{1}=B_{2}\right]=1.
\end{equation}
But we know from (\ref{eq:distributions}) for context 22 that the
probability of $A_{2}=B_{2}$ is zero, not 1. We have run into a contradiction. 

A contradiction always means that some of the assumptions made in
the process of reasoning, perhaps unawares, are wrong. The first impulse
one might have is to declare that the random variables with the distributions
(\ref{eq:distributions}) are impossible, but this can easily be dismissed.
One way to do this is to note that the same reasoning with the same
contradiction at the end can be obtained with distributions that are
empirically observed and codified by a well-established theory. This
is the argument chosen by Mermin, who uses distributions that are
predicted by quantum mechanics for a certain choice of the four settings
(directions in which spins are measured) in the standard EPR/Bohm
experiment with spin-$\nicefrac{1}{2}$ particles \cite{CHSH1969}
(see footnote \ref{fn:Mermin-does-mention}). Of course, one can always
challenge the validity of quantum mechanics and the veracity of the
experiments corroborating its predictions (which one would have to
do if the contradiction we arrived at could not be dissolved by any
other means). A much better argument therefore would be to simply
note that the random variables with distributions (\ref{eq:distributions})
exist mathematically, as appropriately chosen functions on certain
probability spaces. We will get to this later, however.

Assuming we are satisfied there is nothing wrong with our distributions,
where else can one seek the cause of the contradiction we derived?
One might try to deny the possibility that Bob's setting have no influence
on Alice's measurements (which probably remains the most commonly
held interpretation of nonlocality among non-physicists). However,
this would be wrong (``disquieting,'' Mermin says) in view of what
physics says about information propagation: e.g., it is ruled out
if Alice's and Bob's measurements are spacelike separated. Note that
Alice has no means to infer Bob's setting because in contexts $i1$
and $i2$ the random variables representing the outcomes of her measurements
have indistinguishable distributions ($+1$ and $-1$ occurring with
equal probabilities). Therefore the hypothetical ways in which Bob's
setting would influence Alice's measurements would have to be contrived
to remain hidden, in addition to contradicting physical theory.

Mermin too dismisses the ``hidden action at a distance'' resolution
of the contradiction he derives, and he suggests that the culprit
is SBP.
\begin{quote}
{\small{}Many people want to conclude from this {[}the contradiction
-- E.D.{]} that what happens at A does depend on how the switch is
set at B, which is disquieting in view of the absence of any connections
between the detectors. The conclusion can be avoided, if one renounces
the Strong Baseball Principle, maintaining that indeed what happens
at A does not depend on how the switch is set at B, but that this
is only to be understood in its statistical sense, and most emphatically
cannot be applied to individual runs of the experiment. To me this
alternative conclusion is every bit as wonderful as the assertion
of mysterious actions at a distance. I find it quite exquisite that,
setting quantum metaphysics entirely aside, one can demonstrate directly
from the data and the assumption that there are no mysterious actions
at a distance, that there is no conceivable way consistently to apply
the Baseball Principle to individual events (p. 49)}.
\end{quote}
While it is quite obvious to me that SBP is wrong, I do not think
Mermin's explanation is sufficiently transparent. Let us try to understand
it. Mermin says ``individual runs'' because in his exposition he
does not even mention random variables, speaking instead of very long
sequences of realizations thereof. If treated informally, this only
obfuscates analysis, and if treated rigorously, complicates it. A
sequence of realizations of a random variables is a random process,
an indexed set of identically distributed random variables. I suggest
therefore that Mermin's transplantation of a sequence of realizations
from one context to another should simply be understood as placing
in these contexts one and the same random variable, the way we have
done this in (\ref{eq:transplanting1}) and (\ref{eq:transplanting2}).
And this is what must not be done, Mermin says based on the contradiction
this led us into, and I think there is no reasonable way to disagree
with this prohibition. Consider any single context, say, 11 in (\ref{eq:context 11}).
The realizations of $A_{1}$ and $B_{1}$ there come in pairs, they
have therefore a well-defined joint distribution. In particular, one
has an opportunity to decide, by looking at a long enough sequence
of the paired realizations, whether they are perfectly correlated.
By contrast, in (\ref{eq:transplanting1}), if one looks at $A_{1}$
in context 11 and $A_{1}$ in context 12, their realizations cannot
co-occur, because context 11 and 12 are mutually exclusive. One has
no non-arbitrary way of pairing a value of $A_{1}$ in context 11
with a value of $A_{1}$ in context 12. There is no meaningful sense
of asking whether they are correlated, perfectly or otherwise. But
then it means that we have made a mistake by denoting them by the
same symbol: $A_{1}$ is a single random variable, even if mentioned
many times, and $A_{1}=A_{1}$ holds with probability 1. In contexts
11 and 12 therefore we have two different random variables with one
and the same distribution. This is, I suggest, how one could understand
Mermin's assertion ``that indeed what happens at A does not depend
on how the switch is set at B, but that this is only to be understood
in its statistical sense.''

How should we amend the representations (\ref{eq:transplanting1})
and (\ref{eq:transplanting2}) to avoid contradiction? The answer
is simple: we should use different symbols for $A_{1}$ in context
11 and $A_{1}$ in context 12, e.g., denote them by $A_{1}^{11}$
and $A_{1}^{12}$, respectively (and analogously for other pairs of
random variables transplanted from one context to another by SBP).\footnote{This notation for the random variables has redundancy in it: the subscripts
can be recovered from the superscripts (contexts) and the symbol ($A$
or $B$) used for the random variable. Thus, $A_{1}^{11}$ could simply
be written as $A^{11}$, and $B_{2}^{12}$ as $B^{12}$. I leave the
notation redundant by a deliberate choice, however, to make the structure
of the random variables maximally transparent.} One can write $A_{1}^{11}\sim A_{1}^{12}$, where $\sim$ means ``is
distributed as'', but statements like $A_{1}^{11}=A_{1}^{12}$, $A_{1}^{11}\not=A_{1}^{12}$,
$A_{1}^{11}+A_{1}^{12}=2$, etc. are all void of meaning because $A_{1}^{11}$
and $A_{1}^{12}$ posses no joint distribution. The system of random
variables in our example can now be presented as follows:

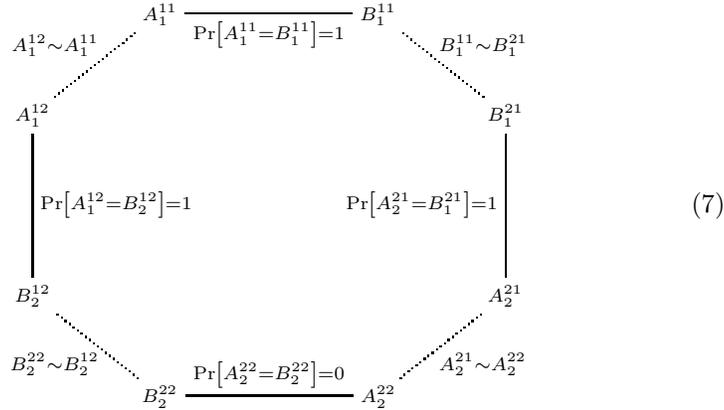
\begin{equation}
\vcenter{\xymatrix@C=1cm{ & {\scriptstyle A_{1}^{11}}\ar@{-}[rr]_{\Pr\left[A_{1}^{11}=B_{1}^{11}\right]=1} &  & {\scriptstyle B_{1}^{11}}\ar@{.}[dr]^{B_{1}^{11}\sim B_{1}^{21}}\\
{\scriptstyle A_{1}^{12}}\ar@{.}[ur]^{A_{1}^{12}\sim A_{1}^{11}} &  &  &  & {\scriptstyle B_{1}^{21}}\ar@{-}[dd]_{\Pr\left[A_{2}^{21}=B_{1}^{21}\right]=1}\\
\\
{\scriptstyle B_{2}^{12}}\ar@{-}[uu]_{\Pr\left[A_{1}^{12}=B_{2}^{12}\right]=1} &  &  &  & {\scriptstyle A_{2}^{21}}\ar@{.}[dl]^{A_{2}^{21}\sim A_{2}^{22}}\\
 & {\scriptstyle B_{2}^{22}}\ar@{.}[lu]^{B_{2}^{22}\sim B_{2}^{12}} &  & {\scriptstyle A_{2}^{22}}\ar@{-}[ll]_{\Pr\left[A_{2}^{22}=B_{2}^{22}\right]=0}
}
}\label{eq:diagram}
\end{equation}
We are no longer driven into contradiction: the value of $\Pr\left[A_{2}^{22}=B_{2}^{22}\right]$
can in no way be inferred from other components of this diagram, because
none of them contains the variables $A_{2}^{22}$ and $B_{2}^{22}$.
If my interpretation of Mermin's conclusion is deemed reasonable,
his was a valuable observation for the 1980s, although I doubt it
could be well understood the way it was formulated. 

To generalize, a random variable within a system of random variables
is identified not only by what it measures (e.g., $A_{1}^{11}$ and
$A_{1}^{12}$ both measure the same property, say, the spin of a particle
in Alice's direction 1) but also by the context in which it is recorded
(here, by the directions chosen by both Alice and Bob for their measurements).
Let me dispel two possible objections to this general statement. 

One is that we still can write $A_{1}$ for both $A_{1}^{11}$ and
$A_{1}^{12}$ but keep in mind that we deal with two different sequences
of realizations of $A_{1}$. Indeed, one might argue, there is nothing
wrong in saying that one records a random variable today, and then
the same random variable is recorded tomorrow. The response to this
argument is that it is acceptable only if one is allowed to be informal,
hoping this will not lead to confusion. A rigorous treatment of random
variables requires that whenever one attaches different contexts to
them (in this example, day of the measurement, today or tomorrow),
one deals with different random variables. After all, to say ``$A_{1}$
today'' and ``$A_{1}$ tomorrow'' means to denote them differently,
albeit sloppily. They are (perhaps) identically distributed, but they
are distinct and have no joint distribution. Recall that the rigorous
definition of a sequence of realizations of a random variable $R$
(a sample of its values) is the sequence of different random variables,
$R_{1},R_{2},\ldots$, each of which is distributed as $R$. The fact
that in most applications they are also assumed to be independent
is more subtle, and its mathematical meaning is captured through the
notion of couplings that we will discuss later. 

Another, often heard objection is that by saying that $A_{1}^{11}$
and $A_{1}^{12}$ are different random variables, one somehow admits
that something in the contexts \emph{causes} $A_{1}^{11}$ to \emph{transform}
into $A_{1}^{12}$ as context 11 is replaced with context 12. So in
the EPR/Bohm scenario, one might argue, we still have some kind of
an action at a distance. This objection is merely a play on the words
``causes'' and ``transforms.'' If $A_{1}^{11}$ and $A_{1}^{12}$
are identically distributed, Alice has no means to distinguish them.
Which means that no information, no action is transferred from Bob's
setting to Alice's measurements. $A_{1}^{11}$ and $A_{1}^{12}$ are
different random variables only for someone who, like Mermin, gets
information from both Alice and Bob, both about their settings chosen
and outcomes obtained. What changes when context 11 is replaced with
context 12 is the relation between Alice's measurements and Bob's
measurements, and this, because of the fundamentally relational nature
of random variables (as explained below), means that Mermin knows
that in contexts 11 and 12 Alice deals with two different random variables. 

Having dealt with these objections, can we say that with my interpretation
of Mermin's analysis we, at least informally (because a more formal
treatment is to follow), have explained the strangeness of contextuality
in terms of random variables? The answer is, we have not. In fact,
unexpected as this might come, Mermin's conclusion in my interpretation
equally applies to any and all systems of random variables, contextual
or not. Consider, e.g., the following modification of the distributions
(\ref{eq:distributions}): 
\begin{equation}
\begin{array}{c}
\begin{array}{c|c|c|c}
\textnormal{contexts 11, 12, 21, 22} & \begin{array}{c}
\textnormal{Bob's}\\
\textnormal{outcome}
\end{array}=+1 & \begin{array}{c}
\textnormal{Bob's}\\
\textnormal{outcome}
\end{array}=-1\\
\hline \textnormal{Alice's outcome}=+1 & \nicefrac{1}{2} & 0 & \nicefrac{1}{2}\\
\hline \textnormal{Alice's outcome}=-1 & 0 & \nicefrac{1}{2} & \nicefrac{1}{2}\\
\hline  & \nicefrac{1}{2} & \nicefrac{1}{2}
\end{array}\end{array}.\label{eq:trivial distribution}
\end{equation}
This system is clearly noncontextual. It can be viewed as describing
a single pair of perfectly correlated random variables, with setting
choices being fake. It is still true, however, that the contexts 11
and 12 are mutually exclusive, and insofar as these settings are not
ignored, Alice's measurements in these contexts must be represented
by different (though identically distributed) random variables that
have no joint distribution. The diagram representing this situation
is identical to (\ref{eq:diagram}), except $\Pr\left[A_{2}^{22}=B_{2}^{22}\right]$
is now 1 rather than zero. This makes no difference for how one treats
the random variables because in both cases $\Pr\left[A_{2}^{22}=B_{2}^{22}\right]$
is completely unrelated to other elements of the diagrams.

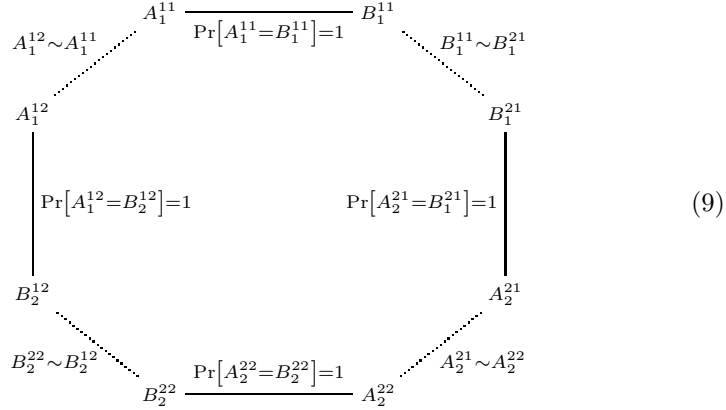
\begin{equation}
\vcenter{\xymatrix@C=1cm{ & {\scriptstyle A_{1}^{11}}\ar@{-}[rr]_{\Pr\left[A_{1}^{11}=B_{1}^{11}\right]=1} &  & {\scriptstyle B_{1}^{11}}\ar@{.}[dr]^{B_{1}^{11}\sim B_{1}^{21}}\\
{\scriptstyle A_{1}^{12}}\ar@{.}[ur]^{A_{1}^{12}\sim A_{1}^{11}} &  &  &  & {\scriptstyle B_{1}^{21}}\ar@{-}[dd]_{\Pr\left[A_{2}^{21}=B_{1}^{21}\right]=1}\\
\\
{\scriptstyle B_{2}^{12}}\ar@{-}[uu]_{\Pr\left[A_{1}^{12}=B_{2}^{12}\right]=1} &  &  &  & {\scriptstyle A_{2}^{21}}\ar@{.}[dl]^{A_{2}^{21}\sim A_{2}^{22}}\\
 & {\scriptstyle B_{2}^{22}}\ar@{.}[lu]^{B_{2}^{22}\sim B_{2}^{12}} &  & {\scriptstyle A_{2}^{22}}\ar@{-}[ll]_{\Pr\left[A_{2}^{22}=B_{2}^{22}\right]=1}
}
}\label{eq:diagramfake}
\end{equation}
Somewhat paradoxically, therefore, having resolved the contradiction
brought in by SBP, Mermin (or at least my interpretation of his analysis)
loses the distinction between contextual and noncontextual systems.
There is no way, e.g., to derive Bell-type inequalities for systems
like (\ref{eq:diagram}) and (\ref{eq:diagramfake}), because the
joint distributions involved of the four pairs of random variables
\begin{equation}
\left(A_{1}^{11},B_{1}^{11}\right),\left(A_{2}^{21},B_{1}^{21}\right),\left(A_{2}^{22},B_{2}^{22}\right),\left(A_{1}^{12},B_{2}^{12}\right),
\end{equation}
are logically unrelated to each other. 

Some researchers derive from this that the notion of contextuality
is flawed. In particular, Bell-type inequalities, according to this
view, are simply invalid, based on the mistake of following SBP. It
seems that this is also Andrei Khrennikov's view \cite{Avisetal.2009},
although his implementation of the context-dependence of random variables
is different from the one presented here \cite{DzhKuj2014}. 

While one is free not to make distinctions one finds uninteresting,
I find this position less than constructive. It is true that the context-indexing
of random variables precludes the naive notion of contextuality, but
what one better get rid of is the naivety rather than the notion.
As it turns out, there is a conceptual and mathematical tool that
enables us to readily distinguish systems like (\ref{eq:diagram})
from those like (\ref{eq:diagramfake}). As a bonus, this tool, while
incompatible with SBP, justifies and formalizes the counterfactual
reasoning on which SBP is based. The questions like ``what would
the outcome of one's measurement be if it were made in a context other
than the one in which it is made'' translate into rigorous and non-controversial
mathematical problems. 

\section{Sample spaces and couplings}

The mathematical tool in question is (probabilistic) \emph{coupling}.
Before introducing it, however, let us make sure we understand why
two random variables in different contexts do not have a joint distribution
(from which it follows also that they can never be one and the same
random variable).

All random variables in this paper are assumed to be \emph{dichotomous},\footnote{In my favorite theory of contextuality, confining the consideration
to dichotomous variables is not a loss of generality, because all
random variables within a system have to be replaced by sets of jointly
distributed dichotomous variables before the system can be subjected
to contextuality analysis \cite{DzhKuj2017,DzhCerKuj2017}.} because of which a random variable is defined as a function $X:S\rightarrow\left\{ -1,1\right\} $,
with the following properties: 
\begin{enumerate}
\item $S$ belongs to a probability space $\left(S,\Sigma,\mu\right)$,
where $\Sigma$ is a sigma algebra of subsets of $S$, and $\mu$
a probability measure $\Sigma\rightarrow\left[0,1\right]$;
\item $X^{-1}\left(\left\{ 1\right\} \right)\in\Sigma$, and $\Pr\left[X=1\right]=\mu\left(X^{-1}\left(\left\{ 1\right\} \right)\right)$;
$X^{-1}\left(\left\{ -1\right\} \right)\in\Sigma$, and $\Pr\left[X=-1\right]=\mu\left(X^{-1}\left(\left\{ -1\right\} \right)\right)$;
\item $\Pr\left[X=1\right]+\Pr\left[X=-1\right]=1$.
\end{enumerate}
The set $S$ is often called a \emph{sample space}, but I prefer to
use this term for the probability space $\left(S,\Sigma,\mu\right)$.
Random variables $X$ and $Y$ are jointly distributed if and only
if they are functions on the same sample space. If they are, then
\begin{equation}
\Pr\left[X=1,Y=1\right]=\mu\left(X^{-1}\left(\left\{ 1\right\} \right)\cap Y^{-1}\left(\left\{ 1\right\} \right)\right).
\end{equation}
Realizations of $X$ and $Y$ are then defined in pairs. If they are
not on the same sample space, $\Pr\left[X=1,Y=1\right]$ is undefined,
and no pairing scheme for their realizations exists. 

We see that, by definition, to construct a set of jointly distributed
random variables means to specify a sample space and define these
random variables as functions on this sample space. What usually remains
unclear to a student of these textbook definitions is the nature of
a sample space. What is it and how can it be (re)constructed? The
answer to this question is so simple that it can be surprising. Let
us discuss this answer in detail, using a system of the same format
as above, but this time with more arbitrary joint distributions in
the four contexts: for $i\in\left\{ 1,2\right\} $ and $j\in\left\{ 1,2\right\} $,
\begin{equation}
\begin{array}{c}
\begin{array}{c|c|c|c}
\textnormal{context }ij & B_{j}^{ij}=+1 & B_{j}^{ij}=-1\\
\hline A_{i}^{ij}=+1 & r_{ij} & p_{i}-r_{ij} & p_{i}\\
\hline A_{i}^{ij}=-1 & q_{j}-r_{ij} & 1-p_{i}-q_{j}+r_{ij} & 1-p_{i}\\
\hline  & q_{j} & 1-q_{j}
\end{array}.\end{array}\label{eq:distributions general1}
\end{equation}
The only constraint imposed by my notation here is that the distribution
of $A_{i}^{ij}$ is the same for $j=1$ and $j=2$ (in both cases
$\Pr\left[A_{i}^{ij}=1\right]=p_{i}$), and analogously for the distribution
of $B_{j}^{ij}$. This property of a system of random variables is
called \emph{consistent connectedness}, or \emph{no-disturbance}.
This is precisely the property that guarantees that Alice has no way
of distinguishing $A_{i}^{i1}$ and $A_{i}^{i2}$, the difference
being only available to someone who receives information about both
Alice's and Bob's settings and outcomes. 

We assume, of course, that the distribution in (\ref{eq:distributions general1})
is well-defined, that is, all probabilities shown are numbers between
0 and 1. As it turns out, this is all one needs to say that the random
variables $A_{i}^{ij}$ and $B_{j}^{ij}$ with this joint distribution
exist as mathematical objects. Indeed, consider the following probability
space $\left(S,\Sigma,\mu_{ij}\right)$ for context $ij$: $S=\left\{ a,b,c,d\right\} $,
$\Sigma=2^{S}$, and $\mu_{ij}$ is defined by the probability mass
function
\begin{equation}
\begin{array}{r||c|c|c|c}
x= & a & b & c & d\\
\hline \mu_{ij}\left(\left\{ x\right\} \right)= & r_{ij} & p_{i}-r_{ij} & q_{j}-r_{ij} & 1-p_{i}-q_{j}+r_{ij}
\end{array}.
\end{equation}
The random variables are now defined as the functions
\begin{equation}
A_{i}^{ij}\left(x\right)=\left\{ \begin{array}{cc}
+1 & \textnormal{if }x\in\left\{ a,b\right\} \\
-1 & \textnormal{if }x\in\left\{ c,d\right\} 
\end{array}\right.,B_{j}^{ij}\left(x\right)=\left\{ \begin{array}{cc}
+1 & \textnormal{if }x\in\left\{ a,c\right\} \\
-1 & \textnormal{if }x\in\left\{ b,d\right\} 
\end{array}\right..
\end{equation}
The resulting system of random variables can be presented in the form
of the following \emph{content-context matrix}:
\begin{equation}
\begin{array}{|c|c||c|c||c|}
\hline A_{1}^{11} &  & B_{1}^{11} &  & \textnormal{context }11\\
\hline A_{1}^{12} &  &  & B_{2}^{12} & \textnormal{context }12\\
\hline  & A_{2}^{21} & B_{1}^{21} &  & \textnormal{context }21\\
\hline  & A_{2}^{22} &  & B_{2}^{22} & \textnormal{context }22\\
\hline\hline \textnormal{Alice's }1 & \textnormal{Alice's }2 & \textnormal{Bob's }1 & \textnormal{Bob's }2 & \textnormal{Alice-Bob system}
\\\hline \end{array}.\label{eq:systemAB}
\end{equation}
The term \emph{content} (of a random variable) refers to that which
the random variable is measuring, or settings from which the measured
property can be deduced. In the matrix above, the contents are listed
at the bottom.

Clearly, up to the labeling of the values of $S$, this construction
is unique. Moreover, any other sample space $\left(S_{ij}',\Sigma_{ij}',\mu_{ij}'\right)$
on which $A_{i}^{ij}$ and $B_{j}^{ij}$ can be defined is reducible
to this $\left(S,\Sigma,\mu_{ij}\right)$, in the following sense:
denoting by $X_{a}$ the pre-image of $\left[A_{i}^{ij}=+1,B_{j}^{ij}=+1\right]$
in $S_{ij}'$, by $X_{b}$ the pre-image of $\left[A_{i}^{ij}=+1,B_{j}^{ij}=-1\right]$,
etc., one can map $X_{a}$ into $a$, $X_{b}$ into $b$, and so on,
to define $A_{i}^{ij}$ and $B_{j}^{ij}$ as functions on $\left(S,\Sigma,\mu_{ij}\right)$.
The latter therefore is the most economic sample space possible. The
general logic of the construction should be clear. Whenever a joint
distribution of hypothetical random variables is well-defined, these
random variables exist as functions defined on a sample space, and
the most economic version of the latter can be uniquely constructed
from the joint distribution. There is never a situation in which one
can say that random variables with a given joint distribution do not
exist (provided no \emph{a priori} constraints are imposed on their
sample space). 

It is also clear from this construction why $A^{i1}$ and $A^{i2}$
are distinct random variables even if they are identically distributed.
They are defined on different sample spaces: even if one chooses to
denote the elements of the sample set in the same way, $\left\{ a,b,c,d\right\} $,
the respective measures $\mu_{i1}$ and $\mu_{i2}$ are as distinct
as are the joint distributions in contexts $i1$ and $i2$ in (\ref{eq:distributions general1}). 

In the previous section I mentioned ``the fundamentally relational
nature of random variables,'' because of which the identity of random
variables cannot be the same in different contexts. One can see now
that this expression has a precise mathematical meaning: the sample
space on which a given random variable is defined is determined by
its joint distribution with all other random variables in the same
context. A reasonable analogy is provided by a set of points in a
metric space without coordinates. Each point is characterized by its
distances to the rest of the points, so moving even one of the latter
changes the point in question instantly. No spooky transfer of information
is involved, these are changes that occur by definition. Of course,
like all analogies, this one also has its drawbacks. In particular,
it is possible to say that ``this point'' (one pointed at) changes
its identity when other points change their positions. In a system
of random variables one can only say that a random variable in one
context is different from a random variable that measures the same
thing in a different context. However, it seems to me that the analogy
with distances does a very good job in dispelling remnants of mystery
in the term ``nonlocality.''

Let me now introduce the conceptual tool that would allow us to speak
of contextual and noncontextual systems. A \emph{coupling} of several
random variables $X_{1},\ldots,X_{n}$ is a set of jointly distributed
variables $Y_{1},\ldots,Y_{n}$ such that $Y_{i}\sim X_{i}$ for $i=1,\ldots,n$.
Note that $X_{1},\ldots,X_{n}$ need not be jointly distributed, and
in fact in all applications we are interested in, they are not. In
other words, each of $X_{i}$ is defined on its own sample space,
whereas all $Y$'s are defined on yet another sample space. Using
the term ``s\emph{tochastically unrelated}'' for random variables
no two of which possess a joint distribution, $X_{1},\ldots,X_{n}$,
and $\left(Y_{1},\ldots,Y_{n}\right)$ viewed as a single random variable,
are stochastically unrelated. In relation to our discussion of Mermin's
SBP, couplings can be thought of as answers to the counterfactual
question ``How could these random variables be jointly distributed
if they were jointly distributed?''

Any set $X_{1},\ldots,X_{n}$ of random variables has a coupling,
and generally it has an infinity of couplings, i.e., infinity of $\left(Y_{1},\ldots,Y_{n}\right)$
with different joint distributions (but the same marginal distributions,
because by definition, $Y_{i}\sim X_{i}$ for $i=1,\ldots,n$).\footnote{One could, obviously, create many copies of $\left(Y_{1},\ldots,Y_{n}\right)$,
identically distributed but defined on different sample spaces. We
should agree therefore that we make no distinction between them: a
coupling is entirely identified by its distribution.} Therefore, to use couplings as a means for categorizing the sets
$X_{1},\ldots,X_{n}$, one should only be interested in whether a
set $X_{1},\ldots,X_{n}$ has a coupling \emph{subject to some specified
constraints} \cite{Dzh2018}, as we will see shortly .

In the Contextuality-by-Default (CbD) theory, the notion of a coupling
is applied to a system of random variables in two ways. We first construct
couplings for all pairs of the same-content random variables. We have
four of them in system (\ref{eq:systemAB}): 
\begin{equation}
\left\{ A_{1}^{11},A_{1}^{12}\right\} ,\left\{ A_{2}^{21},A_{2}^{22}\right\} ,\left\{ B_{1}^{11},B_{1}^{21}\right\} ,\left\{ B_{2}^{12},B_{2}^{22}\right\} .\label{eq:4pairs}
\end{equation}
Denoting a coupling of $\left\{ A_{1}^{11},A_{1}^{12}\right\} $ by
$\left(\tilde{A}_{1}^{11},\tilde{A}_{1}^{12}\right)$, we look among
these couplings for one in which the probability of $\tilde{A}_{1}^{11}=\tilde{A}_{1}^{12}$
is as large as possible, given the individual distribution of $\tilde{A}_{1}^{11}\sim A_{1}^{11}$
and $\tilde{A}_{1}^{12}\sim A_{1}^{12}$. The reason one is interested
in such \emph{maximal couplings} is that the maximal probability in
question is a natural measure of similarity between $A_{1}^{11}$
and $A_{1}^{12}$, when they are taken in isolation from their respective
contexts. Put counterfactually, ``if they were jointly distributed
and no other random variables existed,'' they could coincide as often
as this maximal probability. Suppose we have computed these maximal
probabilities, and
\begin{equation}
\begin{array}{cc}
\max\Pr\left[\tilde{A}_{1}^{11}=\tilde{A}_{1}^{12}\right]=\omega_{1}, & \max\Pr\left[\tilde{A}_{2}^{21}=\tilde{A}_{2}^{22}\right]=\omega_{2},\\
\max\Pr\left[\tilde{B}_{1}^{11}=\tilde{B}_{1}^{21}\right]=\omega_{3}, & \max\Pr\left[\tilde{B}_{2}^{12}=\tilde{B}_{2}^{22}\right]=\omega_{4},
\end{array}\label{eq:maxima}
\end{equation}
where each of the maxima is taken over all possible couplings of the
corresponding pair in (\ref{eq:systemAB}).

We next construct a coupling for the entire system (\ref{eq:systemAB}),
or more precisely, a coupling of the four stochastically unrelated
random variables
\begin{equation}
X_{1}=\left(A_{1}^{11},B_{1}^{11}\right),X_{2}=\left(A_{1}^{12},B_{2}^{12}\right),X_{3}=\left(A_{2}^{21},B_{1}^{21}\right),X_{4}=\left(A_{2}^{22},B_{2}^{22}\right).
\end{equation}
Such a coupling consists of the jointly distributed random variables
\begin{equation}
Y_{1}=\left(\bar{A}_{1}^{11},\bar{B}_{1}^{11}\right),Y_{2}=\left(\bar{A}_{1}^{12},\bar{B}_{2}^{12}\right),Y_{3}=\left(\bar{A}_{2}^{21},\bar{B}_{1}^{21}\right),Y_{4}=\left(\bar{A}_{2}^{22},\bar{B}_{2}^{22}\right),\label{eq:overall coupling}
\end{equation}
with $Y_{i}\sim X_{i}$ ($i=1,\ldots,4$). There are generally an
infinity of such couplings, and in each of them we can compute
\begin{equation}
\begin{array}{cc}
\Pr\left[\bar{A}_{1}^{11}=\bar{A}_{1}^{12}\right]=\omega'_{1}, & \Pr\left[\bar{A}_{2}^{21}=\bar{A}_{2}^{22}\right]=\omega'_{2},\\
\Pr\left[\bar{B}_{1}^{11}=\bar{B}_{1}^{21}\right]=\omega'_{3}, & \Pr\left[\bar{B}_{2}^{12}=\bar{B}_{2}^{22}\right]=\omega'_{4}.
\end{array}\label{eq:rho primes}
\end{equation}
Obviously,
\begin{equation}
\omega'_{1}\leq\omega{}_{1},\omega'_{2}\leq\omega{}_{2},\omega'_{3}\leq\omega{}_{3},\omega'_{4}\leq\omega{}_{4}.
\end{equation}
What we need to determine is whether there is a coupling in which
all these inequalities become equalities, i.e.,
\begin{equation}
\begin{array}{cc}
\Pr\left[\bar{A}_{1}^{11}=\bar{A}_{1}^{12}\right]=\omega{}_{1}, & \Pr\left[\bar{A}_{2}^{21}=\bar{A}_{2}^{22}\right]=\omega{}_{2},\\
\Pr\left[\bar{B}_{1}^{11}=\bar{B}_{1}^{21}\right]=\omega{}_{3}, & \Pr\left[\bar{B}_{2}^{12}=\bar{B}_{2}^{22}\right]=\omega{}_{4}.
\end{array}\label{eq:rhos}
\end{equation}
 In other words, we seek couplings of system (\ref{eq:systemAB})
that preserve both the distributions within the contexts (as they
should by the definition of a coupling) and the similarity values
(\ref{eq:maxima}) between the content-sharing variables. If no such
couplings exist, one can say that the contexts make the content-sharing
random variables to be more dissimilar than they are when they were
taken in isolation. Such a system is called \emph{contextual}. Otherwise,
if such a coupling exists (generally not uniquely), the system is
\emph{noncontextual}.

This simple and, I would argue, highly intuitive logic of (non)contextuality
is sufficient to restore all the contextuality results obtained in
the literature and then go much further, doing so without compromising
the rigorous mathematics of random variables and, in particular, without
falling into the SBP trap. For a consistently connected system, as
in (\ref{eq:distributions general1}), the elements of each pair in
(\ref{eq:4pairs}) are identically distributed, and it is easy to
see that in this case
\begin{equation}
\omega_{1}=\omega_{2}=\omega_{3}=\omega_{4}=1.
\end{equation}
Indeed, the maximal coupling $\left(\tilde{A}_{1}^{11},\tilde{A}_{1}^{12}\right)$
of $\left\{ A_{1}^{11},A_{1}^{12}\right\} $, e.g., has the distribution
\begin{equation}
\begin{array}{c|c|c|c}
\textnormal{content }1\textnormal{ of Alice} & \tilde{A}_{1}^{12}=+1 & \tilde{A}_{1}^{12}=-1\\
\hline \tilde{A}_{1}^{11}=+1 & p_{1} & 0 & p_{1}\\
\hline \tilde{A}_{1}^{11},=-1 & 0 & 1-p_{1} & 1-p_{1}\\
\hline  & p_{1} & 1-p_{1}
\end{array}.
\end{equation}
It follows that in this case we seek couplings $\left(Y_{1},Y_{2},Y_{3},Y_{4}\right)$
in (\ref{eq:overall coupling}) such that $\Pr\left[\bar{A}_{1}^{11}=\bar{A}_{1}^{12}\right]=1$,
$\Pr\left[\bar{B}_{1}^{11}=\bar{B}_{1}^{21}\right]=1$, and so on.
Equivalently, we seek a \emph{reduced coupling} \cite{DzhKuj2016}
in which the variables $\bar{A}_{1}^{11}$and $\bar{A}_{1}^{12}$
can be replaced with a single $\bar{A}_{1}$, the variables $\bar{B}_{1}^{11}$and
$\bar{B}_{1}^{21}$ can be replaced with a single $\bar{B}_{1}$,
etc. 

With this reformulation, one comes as close to the intuition underlying
SBP as it is possible without committing the logical error of the
original SBP. Thus, the system in our opening example, with distributions
(\ref{eq:distributions}), is contextual because otherwise it would
have to have a coupling in which 
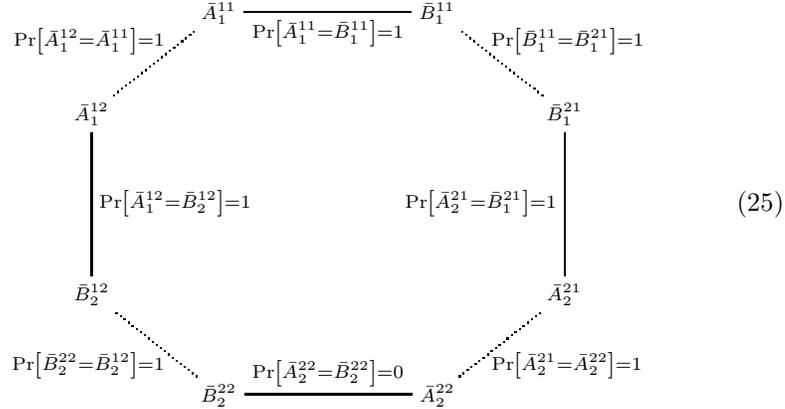
\begin{equation}
\vcenter{\xymatrix@C=1cm{ & {\scriptstyle \bar{A}_{1}^{11}}\ar@{-}[rr]_{\Pr\left[\bar{A}_{1}^{11}=\bar{B}_{1}^{11}\right]=1} &  & {\scriptstyle \bar{B}_{1}^{11}}\ar@{.}[dr]^{\Pr\left[\bar{B}_{1}^{11}=\bar{B}_{1}^{21}\right]=1}\\
{\scriptstyle \bar{A}_{1}^{12}}\ar@{.}[ur]^{\Pr\left[\bar{A}_{1}^{12}=\bar{A}_{1}^{11}\right]=1} &  &  &  & {\scriptstyle \bar{B}_{1}^{21}}\ar@{-}[dd]_{\Pr\left[\bar{A}_{2}^{21}=\bar{B}_{1}^{21}\right]=1}\\
\\
{\scriptstyle \bar{B}_{2}^{12}}\ar@{-}[uu]_{\Pr\left[\bar{A}_{1}^{12}=\bar{B}_{2}^{12}\right]=1} &  &  &  & {\scriptstyle \bar{A}_{2}^{21}}\ar@{.}[dl]^{\Pr\left[\bar{A}_{2}^{21}=\bar{A}_{2}^{22}\right]=1}\\
 & {\scriptstyle \bar{B}_{2}^{22}}\ar@{.}[lu]^{\Pr\left[\bar{B}_{2}^{22}=\bar{B}_{2}^{12}\right]=1} &  & {\scriptstyle \bar{A}_{2}^{22}}\ar@{-}[ll]_{\Pr\left[\bar{A}_{2}^{22}=\bar{B}_{2}^{22}\right]=0}
}
}\label{eq:no such coupling}
\end{equation}
A chain of equalities it contains,
\begin{equation}
\Pr\left[\bar{A}_{1}^{11}=\bar{B}_{1}^{11}\right]=1,\ldots,\Pr\left[\bar{A}_{2}^{22}=\bar{B}_{2}^{22}\right]=0,\ldots,\Pr\left[\bar{A}_{1}^{12}=\bar{A}_{1}^{11}\right]=1,
\end{equation}
is obviously impossible.

As shown in detail in \cite{Dzh2019}, the language of probabilisitic
couplings, when applied to consistently connected systems, allows
one to formalize both the counterfactual formulation of contextuality
and its formulation in terms of the hidden variable models with noncontextual
mapping. I will not be repeating this discussion here. The Bell-type
inequalities for consistently connected systems are derived in essentially
the same way as they are derived traditionally. For instance, system
(\ref{eq:systemAB}) with distributions (\ref{eq:distributions general1})
can be shown to be contextual (i.e., not to have a coupling with the
stipulated properties) if and only if 
\begin{equation}
\max\left(\pm\left\langle A_{1}^{11}B_{1}^{11}\right\rangle \pm\left\langle A_{1}^{12}B_{2}^{12}\right\rangle \pm\left\langle A_{2}^{21}B_{1}^{21}\right\rangle \pm\left\langle A_{2}^{22}B_{2}^{22}\right\rangle \right)>2,
\end{equation}
where the maximum is taken over all choices between $+$ and $-$
in front of each expected value $\left\langle \ldots\right\rangle $
such that the number of the minus signs is odd \cite{DzhKujLar2015,KujDzh2016}.
This is the well-known CHSH inequality, except it is commonly written
as
\begin{equation}
\max\left(\pm\left\langle A_{1}B_{1}\right\rangle \pm\left\langle A_{1}B_{2}\right\rangle \pm\left\langle A_{2}B_{1}\right\rangle \pm\left\langle A_{2}B_{2}\right\rangle \right)>2.
\end{equation}
The latter form, however, is logically flawed, as it employs SBP and
places the same random variable in different contexts. In fact, this
inequality simply cannot be satisfied, because the variables $A_{1},A_{2},B_{1},B_{2}$
in it must be jointly distributed (defined on the same sample space)
by the following diagram:

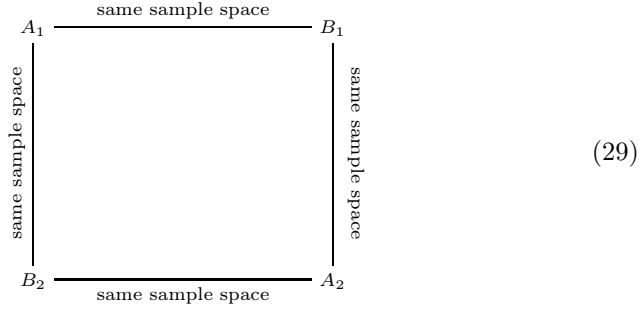
\begin{equation}
\vcenter{\xymatrix@C=1cm{{\scriptstyle A_{1}}\ar@{-}[rrr]^{\textnormal{same sample space}} &  &  & {\scriptstyle B_{1}}\ar@{-}[ddd]^{\textnormal{ \rotatebox{-90}{\textnormal{same sample space}}}}\\
\\
\\
{\scriptstyle B_{2}}\ar@{-}[uuu]^{\textnormal{\textnormal{ \rotatebox{90}{\textnormal{same sample space}}}}} &  &  & {\scriptstyle A_{2}}\ar@{-}[lll]^{\textnormal{same sample space}}
}
}
\end{equation}

\section{Overt influences vs contextuality (or on the magic of words)}

There is one situation in which SBP cannot even be considered. It
is the case of \emph{inconsistently connected systems}, or systems
with disturbance. Using again the system (\ref{eq:systemAB}) as an
example, suppose that the distributions of the random variables are
as follows: for $i\in\left\{ 1,2\right\} $, $j\in\left\{ 1,2\right\} $,
\begin{equation}
\begin{array}{c}
\begin{array}{c|c|c|c}
\textnormal{context }ij & B_{j}^{ij}=+1 & B_{j}^{ij}=-1\\
\hline A_{i}^{ij}=+1 & r_{ij} & p_{ij}-r_{ij} & p_{ij}\\
\hline A_{i}^{ij}=-1 & q_{ij}-r_{ij} & 1-p_{ij}-q_{ij}+r_{ij} & 1-p_{ij}\\
\hline  & q_{ij} & 1-q_{ij}
\end{array}\end{array}.\label{eq:distributions general2}
\end{equation}
The difference between this and (\ref{eq:distributions general1})
is in the marginal distributions: they are no longer necessarily the
same for $A_{i}^{i1}$ and $A_{i}^{i2}$ (generally, $p_{i1}\not=p_{i2}$),
nor are they necessarily the same for $B_{j}^{1j}$ and $B_{j}^{2j}$
(generally, $q_{1j}\not=q_{2j}$).\footnote{The use of ``not necessarily'' here is to indicate that consistently
connected systems, with distributions (\ref{eq:distributions general1}),
are merely a special case of the inconsistently connected ones, with
distributions (\ref{eq:distributions general2}).} Suppose, e.g., that in the EPR/Bohm experiment, Alice's and Bob's
measurements are timelike separated, i.e., transmission of information
from settings of one of them to measurement outcomes of another is
possible. Say, Bob sends certain $\pi$-rays of frequency 1 when he
chooses setting 1, and he sends $\pi$-rays of frequency 2 when he
chooses setting 2, so that the outcomes of Alice's measurements, $A_{i}^{i1}$
and $A_{i}^{i2}$, can be affected by these rays differently. Clearly,
we have here dependence of Alice's measurements not only on her choice
of a setting $i$ but on the entire context $ij$. By definition,
one can speak of \emph{context-dependence} here.

But does this context-dependence necessarily mean that the system
(\ref{eq:systemAB}) with distributions (\ref{eq:distributions general2})
is contextual? Some researchers think that the answer to this question
must be affirmative, unless the distributions are reduced to (\ref{eq:distributions general1}),
in which case a system may be contextual or noncontextual. However,
unless some unknown to me laws compel the meaning of the word ``contextuality''
to be derived from the way it sounds, or its closeness to ``context-dependence,''
this is not the only possible answer. A more constructive approach
would be to consider contextuality as a \emph{form} of context-dependence,
and to ask whether it can be separated from and studied together with
inconsistent connectedness, viewed as another form of context-dependence,
on the level of marginal distributions. 

The definition of contextuality given in the previous section is in
fact formulated for (generally) inconsistently connected systems.
The values of $\omega_{1},\omega_{2},\omega_{3},\omega_{4}$ defined
in (\ref{eq:maxima}) generally are not all equal to 1. Thus, the
maximal couplings of the content-sharing pair $\left\{ A_{1}^{11},A_{1}^{12}\right\} $
now has the distribution
\begin{equation}
\begin{array}{c|c|c|c}
\textnormal{content }1\textnormal{ of Alice} & \tilde{A}_{1}^{12}=+1 & \tilde{A}_{1}^{12}=-1\\
\hline \tilde{A}_{1}^{11}=+1 & \min\left(p_{11},p_{12}\right) & p_{11}-\min\left(p_{11},p_{12}\right) & p_{11}\\
\hline \tilde{A}_{1}^{11},=-1 & p_{12}-\min\left(p_{11},p_{12}\right) & 1-p_{11}-p_{12}+\min\left(p_{11},p_{12}\right) & 1-p_{11}\\
\hline  & p_{12} & 1-p_{12}
\end{array},
\end{equation}
whence 
\begin{equation}
\omega_{1}=\max\Pr\left[\tilde{A}_{1}^{11}=\tilde{A}_{1}^{12}\right]=1-\left|p_{11}-p_{12}\right|.
\end{equation}
Similar formulas hold for other content-sharing pairs. In all other
respects, however, the logic of contextuality remains precisely as
previously described: one seeks an overall coupling of the system
subject to the constraints (\ref{eq:rhos}), and the system is contextual
if and only if no such couplings exist. The interpretation of contextuality
also remains the same as it was for consistently connected systems:
contextuality means that the content-sharing random variable within
their respective contexts (i.e., considered jointly distributed with
other variables) are more dissimilar than when they are isolated from
their contexts. An inconsistently connected system can be contextual
or noncontextual, by precisely the same logic as in the special case
when the system is consistently connected.

This approach is more constructive than simply declaring any inconsistently
connected system contextual, because it provides greater differentiation
among systems of random variables, while properly reducing to special
cases when more restricted definitions apply. One can offer specific
arguments in favor of our definition of contextuality \cite{Dzh2021,DzhKuj2015conversation,DzhKuj2016cc,DzhKuj2018,KujDzh2019,KujDzhLar2016,Dzh2018},
of which I will mention one. First, observe that some contextual systems
are more contextual than others, with respect to the following, intuitively
plausible way of measuring contextuality. Consider the value
\begin{equation}
\begin{array}{l}
\omega'=\max\left(\omega_{1}'+\omega_{2}'+\omega_{3}'+\omega_{4}'\right)\\
=\max\left(\Pr\left[\bar{A}_{1}^{11}=\bar{A}_{1}^{12}\right]+\Pr\left[\bar{A}_{2}^{21}=\bar{A}_{2}^{22}\right]+\Pr\left[\bar{B}_{1}^{11}=\bar{B}_{1}^{21}\right]+\Pr\left[\bar{B}_{2}^{12}=\bar{B}_{2}^{22}\right]\right)
\end{array}
\end{equation}
with the maximum taken over all possible couplings of system (\ref{eq:systemAB}).
This value cannot exceed 
\begin{equation}
\omega_{1}+\omega_{2}+\omega_{3}+\omega_{4}=4-\left|p_{11}-p_{12}\right|-\left|p_{21}-p_{22}\right|-\left|q_{11}-q_{21}\right|-\left|q_{12}-q_{22}\right|,
\end{equation}
because of which the nonnegative quantity
\begin{equation}
\mathsf{CNTX}=\omega_{1}+\omega_{2}+\omega_{3}+\omega_{4}-\omega'
\end{equation}
can be taken for a measure of contextuality. A system is noncontextual
if this quantity is zero. In this paper's opening example, system
(\ref{eq:systemAB}) with distributions (\ref{eq:distributions})
is consistently connected, so $\omega_{1}=\omega_{2}=\omega_{3}=\omega_{4}=1$.
This system is contextual because, as we have seen, it is not possible
for a coupling to satisfy the chain of equalities in (\ref{eq:no such coupling}).
The value of $\mathsf{CNTX}$ for this system can be shown to be 1,
and this can be shown to be the highest possible value of $\mathsf{CNTX}$
across all systems of format (\ref{eq:systemAB}) \cite{KujDzh2016,KujDzh2019,DzhKujCer2020}. 

Now, let us introduce a small disturbance in our example, making the
distributions
\begin{equation}
\begin{array}{c}
\begin{array}{c|c|c|c}
\textnormal{context }22 & B_{2}^{22}=+1 & B_{2}^{22}=-1\\
\hline A_{2}^{22}=+1 & 0 & \nicefrac{1}{2}+\varepsilon & \nicefrac{1}{2}+\varepsilon\\
\hline A_{2}^{22}=-1 & \nicefrac{1}{2}-\varepsilon & 0 & \nicefrac{1}{2}-\varepsilon\\
\hline  & \nicefrac{1}{2}-\varepsilon & \nicefrac{1}{2}+\varepsilon
\end{array}\\
\\
\begin{array}{c|c|c|c}
\textnormal{contexts }ij=11,12,21 & B_{j}^{ij}=+1 & B_{j}^{ij}=-1\\
\hline A_{i}^{ij}=+1 & \nicefrac{1}{2} & 0 & \nicefrac{1}{2}\\
\hline A_{i}^{ij}=-1 & 0 & \nicefrac{1}{2} & \nicefrac{1}{2}\\
\hline  & \nicefrac{1}{2} & \nicefrac{1}{2}
\end{array}
\end{array}.
\end{equation}
Intuition tells us that the degree of contextuality in this system
should be only slightly different from the value of $\mathsf{CNTX}$
in the previous case, for $\varepsilon=0$. And indeed, the degree
of contextuality here is
\begin{equation}
\mathsf{CNTX}=1-2\varepsilon.
\end{equation}
We would not have such a smooth change of the degree of contextuality
with $\varepsilon\rightarrow0$ if we based the contextuality of the
system with $\varepsilon>0$ on the difference of the marginal probabilities
alone.

On the other hand, in our second example, system (\ref{eq:systemAB})
with distributions (\ref{eq:trivial distribution}) is noncontextual,
i.e., $\mathsf{CNTX}=0$. If we introduce the same small perturbation
as above, the distributions will be 
\begin{equation}
\begin{array}{c}
\begin{array}{c|c|c|c}
\textnormal{context }22 & B_{2}^{22}=+1 & B_{2}^{22}=-1\\
\hline A_{2}^{22}=+1 & \nicefrac{1}{2}+\varepsilon & 0 & \nicefrac{1}{2}+\varepsilon\\
\hline A_{2}^{22}=-1 & 0 & \nicefrac{1}{2}-\varepsilon & \nicefrac{1}{2}-\varepsilon\\
\hline  & \nicefrac{1}{2}-\varepsilon & \nicefrac{1}{2}+\varepsilon
\end{array}\\
\\
\begin{array}{c|c|c|c}
\textnormal{contexts }ij=11,12,21 & B_{j}^{ij}=+1 & B_{j}^{ij}=-1\\
\hline A_{i}^{ij}=+1 & \nicefrac{1}{2} & 0 & \nicefrac{1}{2}\\
\hline A_{i}^{ij}=-1 & 0 & \nicefrac{1}{2} & \nicefrac{1}{2}\\
\hline  & \nicefrac{1}{2} & \nicefrac{1}{2}
\end{array}
\end{array}.
\end{equation}
It can be shown that this system remains noncontextual, $\mathsf{CNTX}=0$,
as $\varepsilon$ increases from 0 to $\nicefrac{1}{2}$. Again, this
is what one should expect based on the definition of contextuality.
Zero CNTX at $\varepsilon=0$ means that the system has a coupling
in which the value of $\omega'$ reaches $\omega_{1}+\omega_{2}+\omega_{3}+\omega_{4}$,
which in this case has the maximal possible value, 4. Clearly, this
is even easier to achieve if $\omega_{1}+\omega_{2}+\omega_{3}+\omega_{4}$
has a smaller value, $4-2\varepsilon$. 

As we have seen, an inconsistently connected system can be contextual
or noncontextual, and this lays the ground for a richer classification
of systems than the indiscriminate notion of context-dependence. It
seems reasonable to maintain that being able to make finer differentiations
is always desirable, provided it is done in a principled way. Nevertheless
some researchers keep coming up with the revelatory insight that it
is possible to present both contextuality and inconsistent connectedness
as context-dependence and to refuse to distinguish them. Sometimes
this is presented as the only position consistent with the ``ontological''
(or ``ontic'') models, in which (continuing to use our example)
$A_{i}^{ij}$ and $B_{j}^{ij}$ are presented as functions
\begin{equation}
A_{i}^{ij}=f\left(i,j,\lambda\right),B_{j}^{ij}=g\left(i,j,\lambda\right),\label{eq:stupid}
\end{equation}
where $\lambda$ is some ``hidden'' variable. The term ``ontological/ontic''
is supposed to hint at something happening in reality, as opposed
to purely mathematical descriptions. However, as a general approach,
(\ref{eq:stupid}) is purely descriptive rather than explanatory,
because it is trivially applicable to any system of random variables.
It is in fact nothing more than a mathematically lax version of constructing
an unconstrained overall coupling for system (\ref{eq:systemAB}).
We know that this is always possible. Recall, that to make $\left(A_{i}^{ij},B_{j}^{ij}\right)$
for $i,j\in\left\{ 1,2\right\} $ jointly distributed they have to
be presented as functions on the same sample space. The random variable
$\lambda$ is nothing but the identity function on this sample space.
More rigorously, of course, one has to write
\begin{equation}
\bar{A}_{i}^{ij}=f\left(i,j,\lambda\right),\bar{B}_{j}^{ij}=g\left(i,j,\lambda\right),
\end{equation}
or
\begin{equation}
\left(A_{i}^{ij},B_{j}^{ij}\right)\sim\left(f\left(i,j,\lambda\right),g\left(i,j,\lambda\right)\right),
\end{equation}
because $\left(A_{i}^{ij},B_{j}^{ij}\right)$ for different $i,j$
are stochastically unrelated. 

One source of misunderstanding leading some to considering (\ref{eq:stupid})
as an alternative to CbD is the suggestive terminology I and my colleagues
coined for inconsistent connectedness: we called it (the manifestation
of) \emph{direct influences}, as opposed to contextual influences
\cite{CerDzh2018,CerDzh2019,Basievaetal2019,Dzh2021Abram}. For instance,
the distribution of $A_{i}^{i1}$ may be different from that of $A_{i}^{i2}$
because Bob sends his $\pi$-rays that affect the outcomes of Alice's
measurements. This intuition leads some to point out, as if this were
a discovery of a flaw in CbD, that the $\pi$-rays can also account
for contextuality. One needs nothing but the $\pi$-rays, according
to this reasoning. It is simply that some effects of the $\pi$-rays
are overt, and are reflected in the differences of marginal distributions,
while other effects of $\pi$-rays are hidden, and we call them (mistakenly,
according to this criticism) contextuality \cite{AtmanFilk2019}.
This assertion is being justified, not surprisingly, by the very same
possibility of representing a system by (\ref{eq:stupid}). One can
construct various toy examples to demonstrate this, but the fact remains
that (\ref{eq:stupid}) is applicable universally. As I have mentioned,
it is simply a restatement of the possibility to construct an unconstrained
overall coupling of any system. 

To see that all of this is completely off target, it would suffice
to replace the term \emph{direct influence} with \emph{overt influence}.
In retrospect, this would have been a better term, and I intend to
use it in the future. The criticism in question then would look like
this: CbD distinguishes overt effects (observable on the level of
marginal distributions) and contextual effects in systems of random
variables, while we (the critics) say that some context-dependence
in such systems can be overt and some hidden. This is no more than
a terminological quibble, provided the hidden influences are to be
revealed by means of the CbD-based contextuality analysis. However,
the criticism in question seems to lead its proponents to simply lump
together all context-dependence for systems that are not known to
be consistently connected. If one accepts this position, in physics,
contextuality analysis will be reserved to situations when no physical
transfer of information from Bob's settings to Alice's measurements
(and vice versa) is allowed by laws of physics. If tomorrow the physicists
concluded that superluminal transmission is possible after all, the
EPR/Bohm contextuality would have to be suspect. In systems like the
original Kochen-Specker one \cite{KochenSpecker1967} or KCBS system
\cite{KCBS32008}, where the measurements in each context are made
on the same particle, contextuality is inherently suspect, as there
it hinges on the fact that the current quantum mechanical accounts
of these systems involve no forces or other forms of interference.
In CbD, however, contextuality does not depend on the state of substantive
theories: e.g., the EPR/Bohm system with certain choices of directions
by Alice and Bob is contextual in both contemporary quantum theory
and in Bohmian mechanics, where hidden superluminal transmission is
built in. All of this is discussed and explained in our earlier publications,
e.g. \cite{DzhKuj2015conversation}. Quoting from the latter work, 
\begin{quote}
{\small{}{[}...{]} to defend a definition is a difficult task. A good
definition of a term should be intuitively plausible (although sometimes
one's intuition itself should be ``educated'' to make it plausible),
it should include as special cases all examples and situations that
are traditionally considered to fall within the scope of the term,
it should lead to productive development (to allow one to prove nontrivial
theorems), and have a growing set of applications. I believe contextuality
in the sense of CbD satisfies all these desiderata (p. 14).}{\small\par}
\end{quote}
To summarize:
\begin{enumerate}
\item (Non)contextuality is a property of systems of random variables. It
is a special form of context-dependence, the other form of context-dependence
being inconsistent connectedness.
\item Being a purely mathematical property, (non)contextuality of a system
does not depend on substantive theories of the empirical situations
represented by the system. 
\item The identity of a random variable in a system is determined by its
joint distribution with all other random variables in the same context.
When context changes, a variable measuring some property is instantly
replaced by another random variable measuring the same property (in
the language of CbD, having the same content), or it instantly disappears
(if the property is not measured in the new context). 
\item In particular, if the measurements described by the random variables
in each contexts are separated by spacelike intervals, then the disappearance
or replacement of a random variable by another random variable with
the same content occurs instantly in response to spacelike separated
changes in the context. No action at a distance is involved.
\item The difference between two random variables having the same content
in different contexts is measured by their maximal coupling, and the
system is noncontextual if one of its overall couplings has these
maximal couplings as its marginals. 
\item A contextual system, by contrast, makes the content-sharing random
variables in different contexts more dissimilar than they are in isolation.
\item A system can be contextual or noncontextual irrespective of whether
it is consistently connected.
\end{enumerate}

\end{document}